\documentclass[referee]{aa-5-mpa}
\usepackage{psfig}
\def\aap{A\&A}

\def\ueber#1#2{{\setbox0=\hbox{$#1$}%
  \setbox1=\hbox to\wd0{\hss$ #2$\hss}%
  \offinterlineskip
  \vbox{\box1\box0}}{}}

\def\lesssim{\,\lower 1mm \hbox{\ueber{\sim}{<}}\,}
\def\grsim{\,\lower 1mm \hbox{\ueber{\sim}{>}}\,}

\makeatletter 
\let\@internalcite\cite
\def\cite{\@ifstar{\citeyear}{\citefull}}
\def\citefull{\def\astroncite##1##2{##1 ##2}\@internalcite}
\def\citeyear{\def\astroncite##1##2{##2}\@internalcite}
\def\citeau{\def\astroncite##1##2{##1}\@internalcite}
\def\citen{\def\astroncite##1##2{##1 (##2)}\@internalcite}
\def\possesivcite{\def\astroncite##1##2{##1's (##2)}\@internalcite}
\def\@citex[#1]#2{\if@filesw\immediate\write\@auxout{\string\citation{#2}}\fi
  \def\@citea{}\@cite{\@for\@citeb:=#2\do
    {\@citea\def\@citea{; }\@ifundefined
       {b@\@citeb}{{\bf ?}\@warning
       {Citation `\@citeb' on page \thepage \space undefined}}%
{\csname b@\@citeb\endcsname}}}{#1}}
\def\@cite#1#2{#1\if@tempswa , #2\fi}
\def\@biblabel#1{}
\makeatother

\begin{document}


\title{Atomic diffusion in metal-poor stars}
\subtitle{II.~Predictions for the Spite plateau}

\author{M.~Salaris\inst{1,2} \and A.~Weiss\inst{2,3,4}}

\institute {Astrophysics Research Institute, Liverpool John Moores
University, Twelve Quays House, Egerton Wharf, Birkenhead CH41 1LD, UK
\and
Max-Planck-Institut f\"ur Astrophysik,
Karl-Schwarzschild-Str.~1, 85748 Garching,
Federal Republic of Germany
\and Institute for Advanced Study, Olden Lane, Princeton, USA
\and Princeton University Observatory, Peyton Hall,
Princeton, USA }

\offprints{M.~Salaris; (e-mail: ms@astro.livjm.ac.uk)}

\date{Received; accepted}

\authorrunning{M.~Salaris \& A.~Weiss}
\titlerunning{Diffusion and the Spite plateau}

\abstract{
We have computed a grid of up-to-date stellar evolutionary
models including atomic diffusion, in order to study the
evolution with time of the surface Li
abundance in low-mass metal-poor stars. We discuss in detail the
dependence of the surface Li evolution on the initial metallicity and stellar
mass, and compare the abundances obtained from our models with the
available Li measurements in Pop II stars. While it is widely accepted that
the existence of the Spite Li-plateau for these stars is a strong evidence
that diffusion is inhibited, we show that, when
taking into account observational errors, uncertainties in the Li
abundance determinations, in the 
$T_{\rm eff}$ scale, and in particular the size of the observed samples of stars, 
the Spite plateau and the Li abundances in subgiant branch stars 
can be reproduced also by models including fully efficient diffusion,
provided that the most metal-poor field halo objects are 
between 13.5 and 14~Gyr old. We provide the value of the minimum
number of plateau stars to observe, for discriminating 
between efficient or inhibited diffusion.
{From} our models with diffusion we derive that
the average Li abundance along the Spite plateau is
about a factor of 2 lower than the primordial one.
As a consequence, the derived primordial Li abundance is consistent
with a high helium and low deuterium Big Bang Nucleosynthesis; 
this implies a high cosmological baryon density as inferred
from the analyses of the cosmic microwave background.
 }

\maketitle
\clearpage

\section{Introduction}

Microscopic diffusion is a fundamental physical mechanism, which in
principle has to be considered when computing stellar models. 
In particular, since the
highly accurate seismic determination of the solar sound speed profile
has become available, diffusion is an integral part of the standard solar
model. Taking diffusion into account, not only the sound speed, but
also the depth of the solar convective zone and the present surface
helium abundance are in very good agreement with observations (see,
for example \cite{rvc:96}; \cite{bpbc:97}). Only in
the region immediately below the convective envelope the theoretical
models deviate from the seismic Sun significantly, indicating that
diffusion might not operate exactly in the way calculated or pointing
to some neglected additional physical process partially counter-working
diffusion (\cite{btz:99}). 

With the Sun as our best laboratory for stellar physics strengthening
the case for the presence of atomic diffusion, it immediately becomes
mandatory to consider it also for other stars. Since diffusion is an
intrinsically slow process operating over cosmic times, long-lived
low-mass stars, in particular those of Population~II, are evidently the
best suited candidates. 

Atomic diffusion affects the
evolution of stars in two ways.
In the deep interior, the sinking of helium towards the core leads to an
effectively faster nuclear aging of the star, thus reducing its
Main Sequence (MS) lifetime. For age determinations of globular clusters
making use of the absolute brightness of the colour-magnitude diagram (CMD)
turn-off (TO), when the cluster distance is known, diffusion eventually leads to 
lower ages; the effect accounts for a reduction of the order of 1~Gyr for
old globular clusters (\cite{cddps:92}; \cite{ccdf:97}). The modified
chemical structure also affects 
later evolutionary stages, such that age determination methods using
differential quantities, for example the brightness difference between
TO and horizontal branch, are less affected (about 0.5~Gyr;
\cite{ccd:98}; \cite{cd:99}). 

In the envelope of the star, diffusion leads to a depletion of the
heavy elements and helium. This has the direct consequence
that the present abundance of metals is lower than the initial one. In
\citen{sgw:00} we discussed in detail the consequences for the
MS fitting method, the resulting distances, and subdwarfs
ages. Furthermore, due to the decreased metals, the colours of the
star change. The stellar radius and thus the effective temperature,
however, also depend on the molecular weight in the star's core, which
is increased in the presence of sedimentation. 
Taking diffusion into account, a much better agreement
between models and observed colours or effective temperature results,
as has been noted by \citeau{lpc:99} (\citeyear{lpc:99}; see also
\cite{mb:99} and \cite{sgw:00}). 
Similarly, an improved isochrone fitting for globular clusters, and consistency
between the age from the cluster CMD and that from its integrated spectrum 
might be obtained (\cite{vsar:2001} for 47Tuc). 

A general prediction of models including atomic diffusion 
is that very metal-poor low-mass stars, with their
high effective temperatures and thus very shallow convective envelopes,
are readily depleted of metals over the MS lifetime (see
the seminal paper by \cite{ddk:90}); this
prediction can be contrasted with observations. In the present paper,
we will therefore confront theoretical models with the observational
result that the $^7$Li abundance in such stars of $T_{\mathrm{
eff}}\grsim 5900\;{\mathrm K}$ forms a well-defined plateau below $\mathrm
{[Fe/H]} \approx -1.5$ (\cite{spsp:82}). In the past, this so-called Li- or
Spite-plateau has been used repeatedly as an, if not {\em the}
argument against the full efficiency of
diffusion (\cite{dd:91}; \cite{vc:98};
\cite{rbdt:96}), because the degree of depletion by diffusion increases with
effective temperature. Stated differently,
the necessity for an additional effect counteracting diffusion was
claimed (\cite{cdp:95}; \cite{vauc:99}). Candidate
processes are rotationally induced mixing
(\citeau{cdp:95}~\citeyear{cdp:95}, \citeyear{cdp:95b}) or a
rather strong mass loss by a stellar wind (\cite{vc:95}).

Although, therefore, the Li-plateau seems to make a strong case
against diffusion, there are a number of issues that warrant a new
look at this problem. First of all, the input physics for the stellar
models has changed a lot in the last few years, and the improved
equations of state, opacities and nuclear reaction rates all affect
the stellar models (see \cite{ccd:98} for an overview). Secondly, and as
a consequence, the deduced ages for the oldest 
clusters and stars has been reduced from 16-18~Gyr to 12-14~Gyr
(\cite{sdw:97}). Thirdly, the observed field stars have a distribution of metallicities
and possibly of ages, while often the comparison is performed by
considering models of different masses all with the same age and metallicity
(as, e.g., in \cite{ddk:90} or \cite{vc:98}).
Finally, it is important to consider, when comparing theory with
observations, not only the observational errors, but also the actual
size of the sample of stars with Li abundance determinations.


The inclusion of diffusion does not only lead to a depletion of the
surface lithium content, but to a second effect which could be
measurable; the diffused Li is stored immediately below the thin
convective envelope (see \cite{vc:98} for illustrative interior
profiles) but remains at temperatures too low for lithium-destruction
by proton capture. After the TO, the deepening convective
envelope at first mixes this surplus to the surface, such that for some time 
subgiants should have Li-abundances somewhat {\em above} the Spite-plateau
level which, in the case of effective diffusion, does not represent
the primordial abundance. 
To recall, in case of no diffusion there is no difference between the
initial Li-abundance and that on the main sequence and subgiant branch.
With further expansion the convective
envelope deepens until regions where the initial lithium was destroyed
are diluting the envelope lithium content and a decrease in the
surface abundance will ensue. \citen{psb:93} have
investigated this effect in detail and compared with observations of
metal-poor subgiant. They indeed found indications of this variation
in lithium abundance as a function of evolutionary state.

In this paper we will therefore investigate theoretical predictions
for the Li-abundance\footnote{Throughout the paper we will indicate
the abundance of Li with [Li], defined as [Li]=12+log[N(Li)/N(H)].
We consider only $^7\mathrm{Li}$ in our theoretical predictions for [Li], 
since the cosmological production of $^6\mathrm{Li}$ is negligible in comparison
with $^7\mathrm{Li}$. Observations, however, determine the sum 
Li=$^7\mathrm{Li}$+$^6\mathrm{Li}$.
We will come back to this point in the discussion at the end of the paper.}
of low-mass metal-poor stars, making use of the
latest stellar models, which will be presented in Section~2. The
calculations were used to prepare data about the lithium abundance as
a function of metallicity, effective temperature and age.
The comparison with the observed Li abundances of plateau 
and subgiant branch stars will be described in Section~3, while
in Section~4 we will discuss the
consequences of our results and, in particular, to which extent the
Spite plateau can be used to make statements about the efficiency
of diffusion operating in old metal-poor stars. 

\section{Calculations and models}

We started out with stellar evolution calculations using the same
input physics as in \citen{sw:98} and \citen{sgw:00}. We computed
$\alpha$-enhanced models (average enhancement by 0.4~dex, as in 
\cite{sw:98}) for initial 
iron contents $\mathrm{[Fe/H]_o} = -3.2,\, -2.6,\, -2.3,\, -1.8$ 
(corresponding to $Z=0.000025,\, 0.0001,\, 0.0002, \, 0.0006$)
and masses between
0.6 and 1.0 $M_{\odot}$, from the pre-MS phase until the beginning of
the Red Giant Branch (RGB).  
The initial He content has been fixed at $Y=0.23+3Z$, and we used 
an initial lithium abundance $\rm [Li]_{\rm o}$=2.50.
Diffusion was included by solving the Burgers equations following
\citen{tbl:94}, and considered for H, He, Li, C, N, O, and Fe. 
The other elements have been assumed to diffuse like Fe.
The rates for the Li destruction reactions are from the NACRE compilation
(\cite{an:99}).

Pre-MS evolution was followed
to assess the possible Li-depletion 
in this phase, although it is
expected to lead to substantial lithium destruction only for stellar
masses below $0.8\,M_\odot$ (\cite{dm:84}; \cite{ddk:90}).  
For these pre-MS calculations, the Garching stellar evolution code was used
(see, e.g., \cite{wsch:2000} and references therein), in
contrast to the variant of the FRANEC program for all other
calculations. Both codes use in most aspects essentially the same and
in some (opacities, diffusion coefficients) identical physics and
produce very similar results, as shown, for example, in
\citen{wsch:2000}. 
The derived Li abundances at the Zero Age Main Sequence where then provided
to the code we have employed in all our previous evolutionary computations.
Selected results about the surface Li-depletion exclusively due to nuclear
burning are summarized in Table~\ref{t:lipms}. 

These numbers can be compared with, for example, the results of
\citen{ddk:90}. We get a definitely higher depletion 
at the lower masses than in these older
calculations (with older opacities and equation of state). For
example (see their Fig.~14), our $0.60\;M_\odot$ model shows strong
depletion already on the pre-MS for all metallicities,
and continuing depletion on the
MS, while the depletion in the Z=0.0001 models by \citen{ddk:90} levels off at
$-$0.4~dex already at the beginning of the MS phase. 
This behaviour is
similar to that found by \citen{vzma:98}, when comparing with their
older models (\cite{dm:84}): the more up-to-date physics leads to a
higher depletion. 

\begin{table}
\caption{Surface Lithium depletion (in~dex) -- due to nuclear burning only -- 
along the pre-MS and MS,
for various selected masses ($M/M_\odot$) and initial
metallicities ($Z$). The four lines for each case correspond to the
depletion after 10 and 100~Myr, 1 and 10~Gyr. For simplicity, we omit
lines of higher age when the lithium abundance does not change any
longer. } 
\protect\label{t:lipms}
\begin{center}	
\begin{tabular}{l|ccc}
$M/M_\odot\;\vert\; Z$ & $2.5\,10^{-5}$ & $2\,10^{-4}$&
$6\,10^{-4}$ \\
\hline
0.60 & -0.75 & -0.63 & -0.42 \\
     & -1.03 & -0.90 & -0.79 \\
     & -1.36 & -0.96 & -0.89 \\
     & -1.83 & -1.83 & -1.76 \\
0.65 & -0.19 & -0.20 & -0.19 \\
     & -0.19 & -0.21 & -0.23 \\
     & -0.19 & -0.22 & -0.25 \\
     & -0.19 & -0.22 & -0.26 \\
0.70 & -0.03 & -0.06 & -0.07 \\
     & -0.04 & -0.06 & -0.07 \\
     & -0.04 & -0.06 & -0.08 \\
0.75 & -0.01 & -0.02 & -0.03 \\
0.80 & -0.00 & -0.01 & -0.01 \\
\hline
\end{tabular}
\end{center}

\end{table}

When comparing our diffusion calculations with the results 
from the scaled-solar models by
\citen{vc:98}, for ages of 12 and 14~Gyr, 
we find good agreement, with large differences only 
for the $M=0.65 M_{\odot}$ models,  which are
mainly due to differences in the pre-MS phase, since for these
masses and ages the depletion due to diffusion from the convective
envelope is only a small contribution.

\begin{table}
\protect\label{t:livc}
\caption{
Lithium depletion, 
given as $\mathrm{Li}(t)/\mathrm{Li}_0$,
in models with diffusion. The third column shows results of
Vauclair \& Charbonnel (1998) 
for $\mathrm{[Fe/H]}=-2 $ (``VC''), the fourth one those
of the present paper at the same 
[Fe/H], and the last one at the same total metallicity, [M/H]. The
comparison is done for three masses (column 1) in common and two ages,
12 and 14~Gyr (column 2).}
\begin{center}
\begin{tabular}{l|l|ccc}
$M/M_\odot$ & $t$ & VC & $\mathrm{[Fe/H]}=-2.$ & $\mathrm{[M/H]}=-2.$ \\
\hline
0.65 & 12 & 0.700 & 0.431 & 0.436 \\
     & 14 & 0.662 & 0.412 & 0.427 \\
0.70 & 12 & 0.652 & 0.695 & 0.667 \\
     & 14 & 0.595 & 0.657 & 0.622 \\
0.75 & 12 & 0.465 & 0.482 & 0.398 \\
     & 14 & 0.384 & 0.365 & 0.245 \\
\hline
\end{tabular}
\end{center}
\end{table}

\begin{figure}
\centerline{\psfig{figure=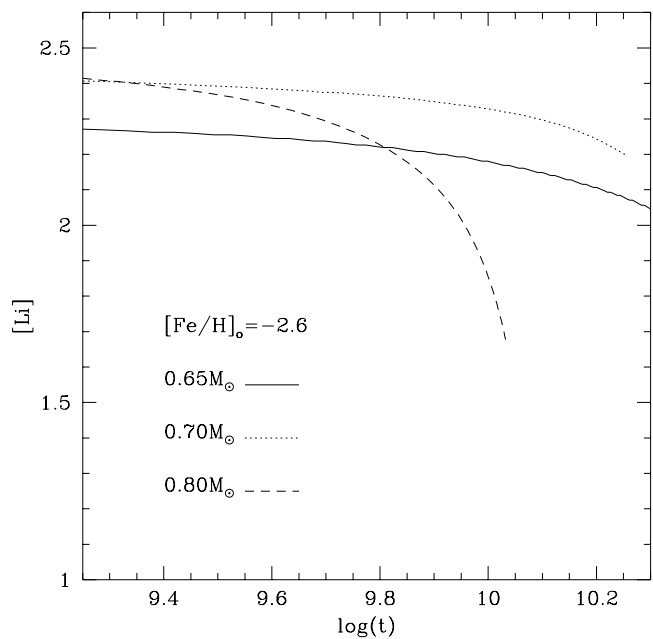,width=9.0cm,clip=}}
\caption[]{Li abundance as a function of the age (t), along the MS,
for three selected masses and initial metallicity $\rm [Fe/H]_{o}=-$2.6.}
\protect\label{trackLi}
\end{figure}

Figure~\ref{trackLi} shows the surface [Li] value as a function of age
for three selected masses (0.65, 0.70 and 0.80 $\rm M_{\odot}$) and
initial metallicity $\rm [Fe/H]_{o}=-$2.6 ($Z=0.0001$), 
along the MS. All the three masses show a steady decrease of [Li] as a
function of time due to the effect of diffusion.
The $0.80 M_{\odot}$ star shows a slightly higher [Li] value during the
first MS stages (due to its lower pre-MS depletion), then, due to the
shrinking of the envelope convection when approaching the TO, [Li]
quickly drops below the values of the 0.65 and $0.70\, M_{\odot}$
objects. The latter two masses, with larger convective envelopes
all along the MS phase, show a much more similar trend of [Li] as a function of
time, the steady [Li] decrease of the $0.70\, M_{\odot}$ star being
however slightly steeper due to its thinner convective envelope; 
the difference in the absolute [Li] values is mainly due to the different
amount of pre-MS depletion.


{From} the computed grid of stellar models we then produced
Li-isochrones (i.e.\ Li-abundances along isochrones as in
\cite{ddk:90}) for different ages and different initial [Fe/H] values.
It is important to remark here that along an isochrone the surface
[Fe/H] is
not constant, again due to the effect of diffusion 
(\cite{mb:99}, \cite{sgw:00}). At a given age of some
Gyr, the surface [Fe/H] is lower than the initial value; it also shows a
decrease towards the TO and than increases along the
Sub Giant Branch (SGB) until it reaches the initial value along the
RGB (see Fig.~\ref{isototal}).
Therefore our isochrones, for ages larger than $\sim$5~Gyr, cover  
a range of surface [Fe/H] between $\sim -$2.0 and $\sim-$3.8.

\begin{figure}
\centerline{\psfig{figure=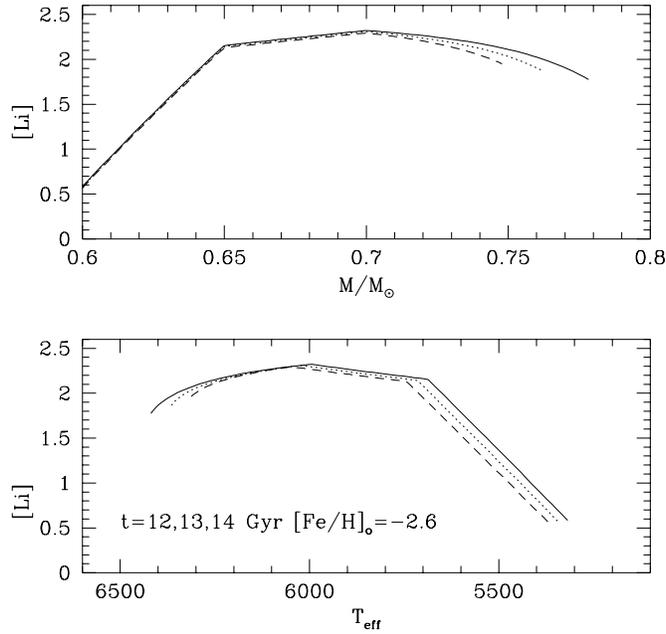,width=9.0cm,clip=}}
\caption[]{(upper panel) Run of [Li] as a function of the stellar mass
for isochrones with initial metallicity $\rm [Fe/H]_{o}=-$2.6
and ages of 12 (solid line), 13 (dotted line) and 14~Gyr (dashed line).
(lower panel) As in the upper panel, but for 
the run of [Li] as a function of $T_{\rm eff}$.}
\protect\label{isoage}
\end{figure}

Figure~\ref{isoage} displays Li-isochrones as a function of 
$T_{\rm eff}$ and mass for ages of 12, 13, 14~Gyr and
initial metallicity $\rm [Fe/H]_{o}=-$2.6, from the unevolved lower main
sequence up to the TO.
The shape of the isochrones looks very similar for the various ages
apart from the different value of $T_{\rm eff}$
at both the TO and at the cool end. 
Li depletion at the coolest end of the isochrones is
dominated by the pre-MS depletion (masses populating these region
range from 0.60 up to $\sim$0.65 $M_{\odot}$), while for $T_{\rm eff}$ 
larger than $\sim$5800~K Li diffusion is mainly responsible for the
surface Li depletion, since pre-MS depletion is negligible. 
The isochrones for $t=12$~Gyr show more depletion at the TO since, in
spite of the shorter time available for diffusion to deplete the surface Li,
TO stars are hotter and have shallower convective regions and therefore the surface Li
depletion is more efficient.
The Li-isochrones show also an extended region, with $T_{\rm eff}$
ranging between $\sim$5800 and 6300~K (the extension depends on the
age), in which the depletion does not change by more than 0.2~dex.
At these ages the depletion at the level of this plateau is of the
order of 0.3~dex. 

\begin{figure}
\centerline{\psfig{figure=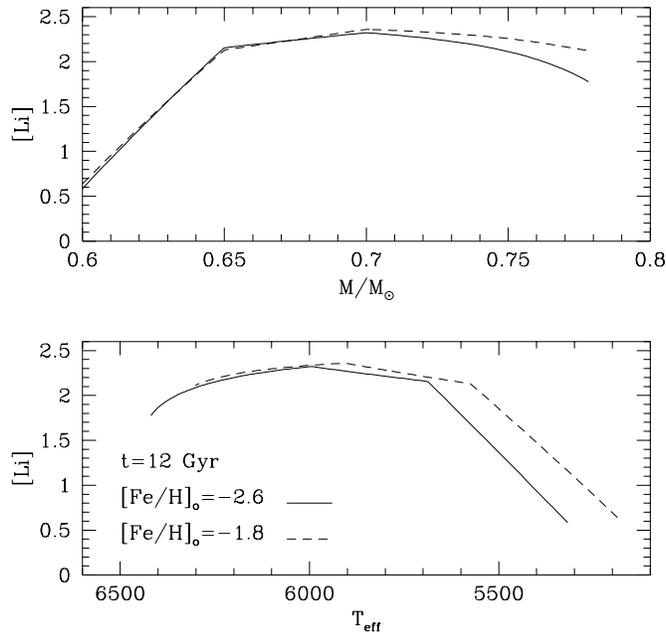,width=9.0cm,clip=}}
\caption[]{As in the previous figure but for isochrones with $t=12$~Gyr
and initial metallicities $\rm [Fe/H]_{o}=-$2.6 and $\rm [Fe/H]_{o}=-$1.8.}
\protect\label{isometal}
\end{figure}

\begin{figure}
\centerline{\psfig{figure=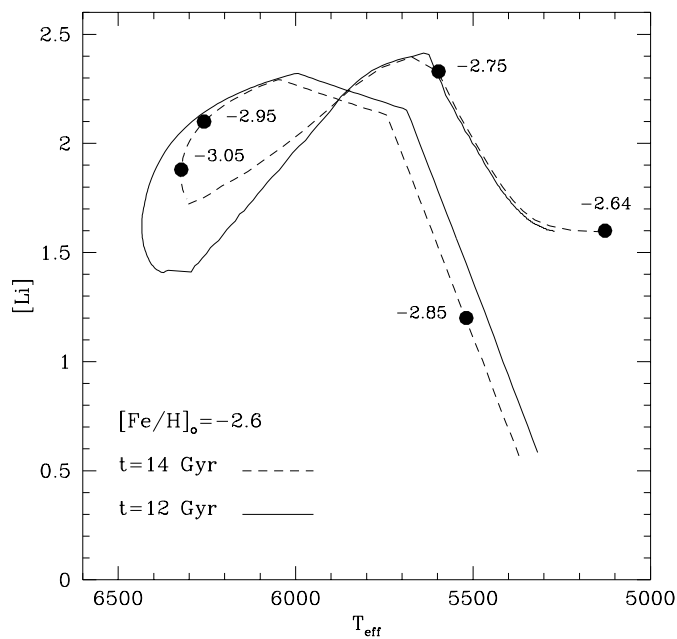,width=9.0cm,clip=}}
\caption[]{[Li] as a function of $T_{\rm eff}$ for isochrones
of 12 and 14~Gyr, with $\rm [Fe/H]_{o}=-$2.6, from the MS to
the lower  RGB. Selected values of [Fe/H] along the
14~Gyr isochrone are also shown.}
\protect\label{isototal}
\end{figure}

In Fig.~\ref{isometal} we compare the [Li] values along two
isochrones with the same age of 12~Gyr and two different initial metallicities.
The shape of the isochrones is again very similar, with the more metal-rich
models being less depleted at the TO (larger convective envelopes)
and in general more Li-rich at a given value of $T_{\rm eff}$.

Finally, in Fig.~\ref{isototal} we display the run of [Li] as a
function of $T_{\rm eff}$ for 12 and 14~Gyr isochrones with 
$\rm [Fe/H]_{o}=-$2.6, from the MS to the base of the
RGB. It is very interesting to notice that after the TO [Li]
continues to decrease for the next 50~K, with a further depletion by
about 0.1~dex. The maximum depletion is a function of age, 
strongly decreasing for increasing age. It also depends on 
metallicity, in the sense that at a given age the maximum depletion
decreases with increasing metallicities. This behaviour is easy to
understand if one recalls that the extension of the convective
envelope in TO stars increases for increasing age at a given 
metallicity, or for increasing metallicity at a given age, 
reducing thus the effect of diffusion on the surface abundances. 

Later on, with decreasing temperature, the [Li] abundance starts to
rise again due to the deepening of the convective region which 
returns to the surface the Li diffused outside the envelope 
along the MS. [Li] reaches a peak
when $T_{\rm eff}\simeq$~5600 -- 5700~K, with an abundance within 0.1~dex
from the initial one; then the Li surface abundance starts to drop, 
since the convective envelope reaches regions where Li had been 
burned due to the higher temperatures of the stellar matter.
We note that the location and level of the Li abundance peak is
fairly independent of age and metallicity (see Fig.~\ref{P93}) and, in
addition, is within 0.1~dex of the primoridial value, independent of
the effectiveness of diffusion. Subgiants at effective temperatures
around $5600\,\mathrm{K}$ are therefore ideally suited for determining
the primordial Li abundance to this accuracy.

\section{Comparison with observations}

In this section we will compare the behaviour of the 
[Li] abundances obtained from our MS and SGB models 
with the observations of metal-poor field and globular cluster stars.
We concentrate our attention on stars 
with [Fe/H]$\le -$2.0, which constitute the largest fraction 
of the currently available samples of PopII Li-plateau stars;
in the last part of this section we will
discuss briefly also the case of stars with [Fe/H] up to $\sim -$1.5, which
is approximately the upper metallicity boundary of the Spite plateau.

\subsection{Field MS stars}

\begin{figure}
\centerline{\psfig{figure=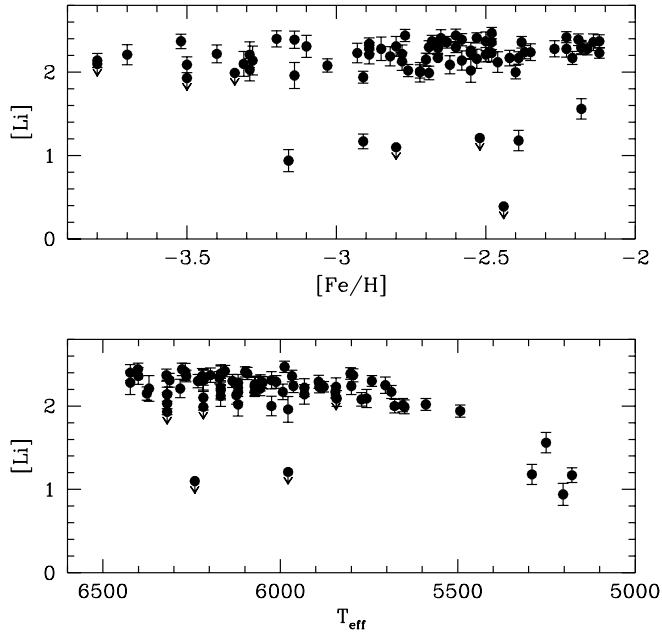,width=9.0cm,clip=}}
\caption[]{Run of [Li] as a function of [Fe/H] (upper panel) and 
$T_{\rm eff}$ (lower panel) from the \citen{t:94} data for MS
stars with [Fe/H]$\leq -$2.0. Upper limit detections are also shown.
For the sake of clarity we did not plot the error bars on the $T_{\rm
eff}$ values (of the order of, typically, 80-100~K, see \cite{t:94} for more 
details).}
\protect\label{LidataT94}
\end{figure}

Observations of Li abundances in metal-poor MS field stars
provide a qualitative picture that is basically consistent among the
different authors, but which can differ a lot in the details
(see, e.g., \cite{t:94}, \cite{rbdt:96}, \cite{sfns:96}, \cite{bm:97},
\cite{rnb:99}, \cite{rkb:01} and references therein).
The main result (as an example we show in Fig.~\ref{LidataT94}
the \cite{t:94} data for the metallicity range spanned by our models)
is that metal-poor MS stars with 
$T_{\rm eff}$ larger than approximately 5800~K show a remarkably
constant [Li] value (Li-plateau), while there is a larger
depletion at lower temperatures, increasing for decreasing temperature.
Moreover, in the plateau region, a handful of stars show a much lower
[Li] than the plateau counterpart (e.g. \cite{t:94}).

The details of this picture, however, differ from author to author. 
The exact $T_{\rm eff}$ location and extent of the plateau varies
slightly; some trend of [Li] with $T_{\rm eff}$ along the plateau was
detected by \citen{t:94} and \citen{rbdt:96}, 
but it is absent in the analyses by \citen{bm:97} and \citen{rnb:99}.
A trend of [Li] with respect to [Fe/H] for plateau stars is claimed by
\citen{t:94}, \citen{rbdt:96}, \citen{rnb:99} and \citen{rkb:01}, 
while it is missing in \citen{bm:97}.
Also the absolute average value of [Li] for plateau stars shows 
differences between different authors.

In addition to the purely observational errors and
differences in the adopted metallicity
scale, there are other sources of uncertainty 
that should be mentioned. The first one is the set of model atmospheres
used in the spectral analyses.
As dicussed by \citen{rbdt:96}, the differences between Li abundances 
derived from different sets of model atmospheres show  
trends with respect to both $T_{\rm eff}$ and [Fe/H].
Very recently \citen{ants:99} discussed the differences in the [Li]
values derived using either 1D, plane
parallel, hydrostatic model atmospheres, like the MARCS (\cite{g:75}
with subsequent updates) and ATLAS9 (\cite{ku:93}) ones used in all recent 
papers on [Li] abundances,
or their 3D hydrodynamical model atmospheres. A remarkable
result is that not only the absolute values of [Li] are changed when
passing from 1D to 3D models, but that also the spread in 
[Li] values is changed. They discuss the case of HD140283 and HD84937
whose [Li]  derived from MARCS models is, respectively, 2.12 and 2.28.
Using their 3D hydrodynamical models, \citen{ants:99} obtain 1.78
and 2.08; the difference between the two values is almost doubled 
with respect to the case of 1D hydrostatic models.
Keeping in mind their cautionary remarks about the
assumption of LTE in their models, it is nevertheless important to
notice that the derived [Li] spread may underestimate
the real variation. 

\begin{figure}
\centerline{\psfig{figure=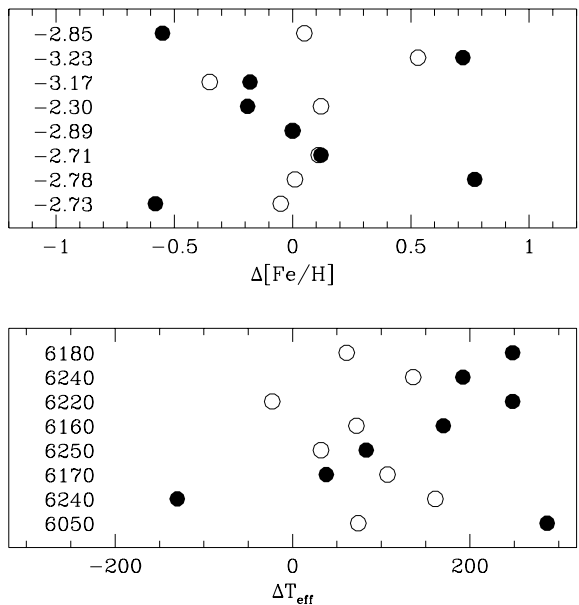,width=9.0cm,clip=}}
\caption[]{Differences in [Fe/H] (upper panel) and
$T_{\rm eff}$ (lower panel) for a sample of 8 plateau stars
with [Fe/H]$\leq -$2.0, in common among two or more of the
investigations mentioned in the text. Each line corresponds to a
different star with [Fe/H] and $T_{\rm eff}$ from \citen{rnb:99}
indicated on the left side. The difference to \citen{bm:97} is indicated
by a filled circle, that to \citen{t:94} by an open one.}
\protect\label{Tdiff}
\end{figure}

Another important source of uncertainty
is the $T_{\rm eff}$ scale itself. Different authors use
temperature scales which often differ not just by a constant
offset, but also show differential trends with respect to the metallicity.
As an example, we show in Fig.~\ref{Tdiff} the values of $T_{\rm eff}$
for 8 plateau stars in common among three of the previously
mentioned investigations.

Clearly 
the spread of the $T_{\rm eff}$ values is not constant and
in some cases quite remarkable. If
one takes into account that a $T_{\rm eff}$ variation by 100~K 
changes the derived [Li] value by about 0.07~dex (\cite{rnb:99}),
it may well be that the current uncertainty about Li 
trends along the plateau is due mainly to uncertainties in the
temperature scale. 
Also the [Fe/H] values adopted in different
investigations show a spread which reaches values larger than 0.5~dex
(Fig.~\ref{Tdiff}).

In order to analyze the predictions of theoretical diffusive models 
regarding the trends of [Li] with respect to $T_{\rm eff}$
and [Fe/H] in metal-poor stars, we resorted to Monte-Carlo (MC) 
simulations using our own isochrones. 
In this way we will be able to take into account the fact that the
subdwarfs with [Li] measurements have a metallicity distribution and
possibly an age distribution. As
we have shown before, the initial metallicity and age alter the
location and shape of
Li--isochrones, and it is important to take these two factors into
account. Moreover, with MC simulations we will be also able to account for 
the actual number of MS stars with [Li] measurements. This point -- as
we will see later on -- is of fundamental importance when trying to
constrain the efficiency of diffusion from Li-plateau stars.
It is important to remark at this stage that, since the observations provide 
contradictory results about the properties of the Spite plateau, 
we can just discuss whether our models with diffusion are able to
reproduce the plateau, and if its derived properties are within the
range allowed by existing data. 

In our simulations we assumed that present [Li] observations of
field MS subdwarfs cover also TO stars.
We prescribed an age--initial metallicity relationship
and, by interpolating among our isochrones, 
we have drawn a large sample of 10000 MS stars, according to a Salpeter
Initial Mass Function, with an initial [Fe/H] distribution adjusted 
(by trial and error) in such a way that the 10000 stars show 
a metallicity distribution in agreement with the typical metallicity
distribution of halo stars as determined by 
\citen{rn:91}.
For each star we determined the value of its mass, bolometric
luminosity, surface [Fe/H] and [Li], and  $T_{\rm eff}$.
We then added to the values of [Fe/H], [Li], and  $T_{\rm eff}$ 
1$\sigma$ random observational errors by, respectively,
0.20~dex, 0.07~dex and 65~K. These observational errors
are approximately an average of the values quoted by \citen{t:94},
\citen{rbdt:96}, \citen{sfns:96}, \citen{bm:97}. 

{From} this sample of 10000 stars representing the halo population we
have randomly drawn 30 smaller samples of 
120 stars each, which represent the MC realizations of the observational
data; about 65 objects populate the plateau region (which we define as being 
the region with $T_{\rm eff}> 5800\,\mathrm{K}$).
This number of plateau stars correponds approximately to the largest sample with [Li]
abundances and [Fe/H]$\leq -$2.0 available to this day (see, e.g.,
\cite{t:94}, \cite{rbdt:96}).
For each of these
samples we studied the distribution of stars in the [Li]-$T_{\rm eff}$
and [Li]-[Fe/H] planes.

\begin{figure}
\centerline{\psfig{figure=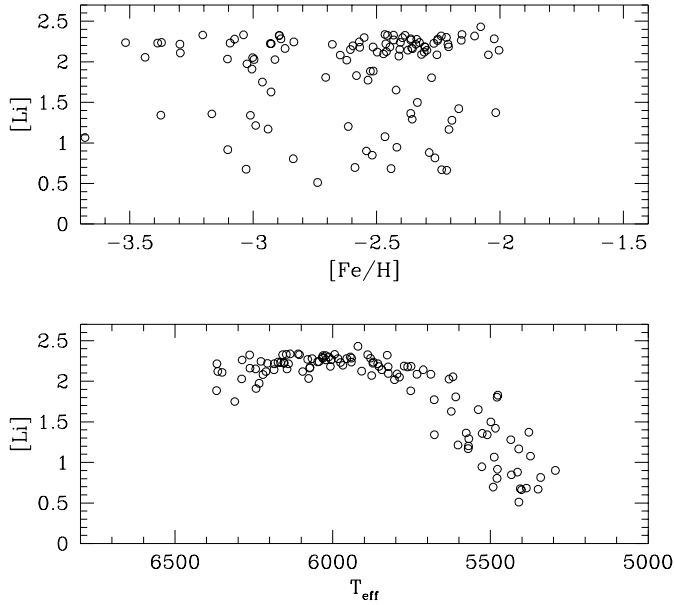,width=9.0cm,clip=}}
\caption[]{ Run of [Li] as a function of [Fe/H] (upper panel) and 
$T_{\rm eff}$ (lower panel) for stars in one of our MC simulations (see
text) considering an age randomly distributed between 13.5 and 14~Gyr.}
\protect\label{sim1}
\end{figure}

We have been able to obtain a plateau-like feature
for $T_{\rm eff}> 5800\,\mathrm{K}$
only when the age of the sample is equal or higher than about 13.5~Gyr. Lower
ages show a pronounced drop of [Li] at the TO which is not observed. 
Above this age, in the limit of the actual
available observational samples and typical errors, the theoretical plateau region 
looks quite similar to the observational counterpart.
In Fig.~\ref{sim1} we display the outcome of one simulation, where the age of the
stars is randomly distributed between 13.5 and 14~Gyr. 
The outcome of this simulation is quite
typical of the ensemble of the 30 samples we have drawn with this age distribution.
The plateau for $T_{\rm eff}$ larger than about
5800~K is clearly visible together with a drop at lower $T_{\rm
eff}$ due to the effect of pre-MS depletion.

\begin{figure}
\centerline{\psfig{figure=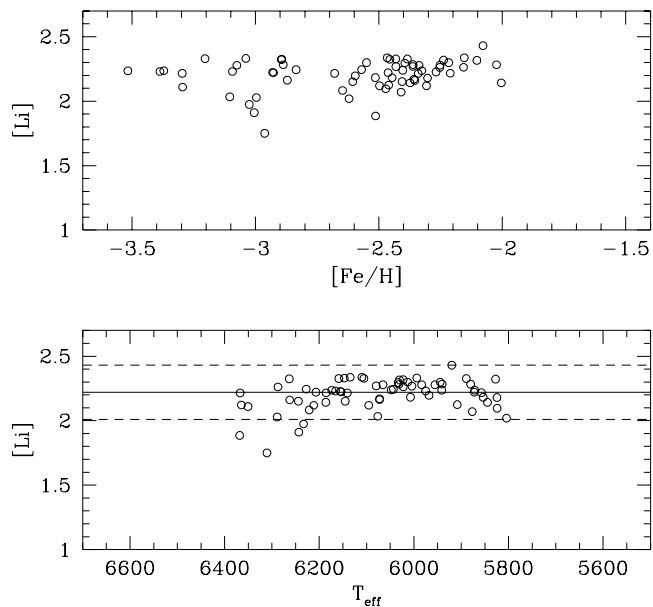,width=9.0cm,clip=}}
\caption[]{ As in the previous figure, but only for plateau stars,
i.e.\ with $T_{\rm eff} > 5800\,\mathrm{K}$
(see text for details).}
\protect\label{sim2}
\end{figure}

Figure~\ref{sim2} displays only the plateau stars from the same simulation.
The solid line in the plane [Li]-$T_{\rm eff}$ shows the average
[Li] value ($\langle$[Li]$\rangle$), while the dashed lines indicate  
$\langle$[Li]$\rangle \pm 3\sigma$, where $\sigma$ is the typical
observational error on [Li] adopted in our simulations (0.07~dex).
Only 4 stars appear to be outside the zone
limited by the dashed lines (with lower [Li] values).
If we take out these 4 stars and recompute $\langle$[Li]$\rangle$, its
value is not affected appreciably and all the remaining stars 
still lie within $3\sigma$ of $\langle$[Li]$\rangle$.
For the stars  within $3\sigma$ of $\langle$[Li]$\rangle$
there is no significant correlation between
[Li] and $T_{\rm eff}$. Stars outside this region 
we will call 'outliers'. In all our simulations the
outliers lie always below $\langle$[Li]$\rangle - 3\sigma$, and they
look like the theoretical counterpart of the observed Li-depleted
stars in the plateau region (see Fig.~\ref{LidataT94},
\cite{t:94} and \cite{rkb:01}). It is important to notice that in the
observational analyses of the plateau, the (few) stars showing a substantial depletion 
with respect to the average plateau value, are not considered in the
derivation of the plateau properties. As we did the same with our
outliers, our selection closely mimicks the observational analyses.

By averaging over the 30 realizations with this same age distribution, we
derived for the plateau stars a [Li] depletion by 0.27~dex and a
derivative ${\rm \Delta[Li]}/{\rm \Delta[Fe/H]}$=0.06. The average
number of outliers (Li-depleted stars) is 6.
This latter number compares well with the actually observed number 
of Li-depleted stars in this same [Fe/H] and $T_{\rm eff}$ range
(see \cite{t:94}). Also the average plateau value [Li]=2.23,
obtained from an initial lithium abundance $\rm [Li]_{\rm o}$=2.50, 
agrees well with the values provided in the literature, which
span a relatively large range 
from $\sim$2.10 (\cite{rnb:99}) up to 
$\sim$2.40 (\cite{gsc:00}).
The slope ${\rm \Delta[Li]}/{\rm \Delta[Fe/H]}$
is halfway between the value 0.118$\pm$0.023 
claimed by \citen{rnb:99}, and the results by 
\citen{bm:97}, who did not find any dependence on [Fe/H] 
for the plateau [Li] values.

The lower $T_{\rm eff}$ limit off the plateau has been, somewhat
arbitrarily, set at 5800~K, which is a typical value as deduced from
the observations. Had we increased it up to 5900~K (again, compatible
with observations), the situation would have changed only slightly,
with just a small decrease of the average [Li] depletion along the
plateau, and a somewhat narrower distribution of the plateau stars
around $\langle$[Li]$\rangle$.

We made simulations also assuming a constant age of, respectively,
13.5 and 14.0~Gyr, and ages between 14 and 15~Gyr; 
the results are substantially the same, in the sense that a
plateau-like feature is always found, without a significant depletion
at its hotter end. We did 
not explore ages larger than about 15~Gyr because of the constraints 
put on the ages of field halo stars by the current estimates of
globular clusters ages (see next section), and also by the fact that
the horizontal extension of the plateau region is very much reduced 
and, even taking into account current uncertainties in the  
$T_{\rm eff}$, it is too much reduced in comparison with
observational data.

\begin{figure}
\centerline{\psfig{figure=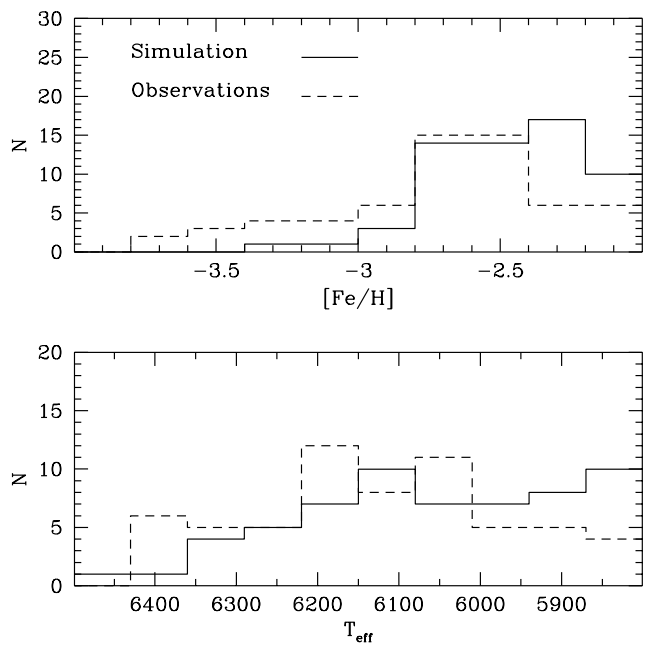,width=9.0cm,clip=}}
\caption[]{Comparison of the number of plateau stars as a function 
of their [Fe/H] (upper panel) 
and $T_{\rm eff}$ (lower panel), between one of our MC samples and
the observational data by \citen{t:94} (see text for details).}
\protect\label{hist1}
\end{figure}

We now investigate whether our assumed metallicity distribution for
halo stars leads to properties of the simulated sample in agreement
with observed ones. 
To this end we compare in Fig.~\ref{hist1}
the distribution of plateau stars 
($T_{\rm eff}> 5800\,\mathrm{K}$) as a function of the actual 
metallicity [Fe/H] and $T_{\rm eff}$ in case of one of our MC samples,
and the data by \citen{t:94} with [Fe/H]$< -$2.0 
(the bin size is equal to 70~K in  
$T_{\rm eff}$ and 0.20~dex in [Fe/H]). The number of plateau 
stars in both cases is the same.

If one takes into account the spread of the empirical
determinations of $T_{\rm eff}$ and [Fe/H] shown in  
Fig.~\ref{Tdiff}, the differences between the distributions of stars
in the observed and synthetic samples are probably not very
significant.
However, as a test, we computed a series of MC simulations 
in the same way as before (ages again distributed
uniformly between 13.5 and 14.0~Gyr), but considering
an actual metallicity distribution biased towards lower metallicities with
respect to the \citen{rn:91} one.
This corresponds, basically, to have synthetic samples of plateau
stars with a higher fraction of very metal-poor stars, in the
assumptions that the observations were biased towards the most metal
poor objects.

The distribution of plateau stars as a function of [Fe/H] and 
$T_{\rm eff}$ for one of these latest simulations is shown in 
Fig.~\ref{hist1}; the agreement between the two distributions is improved.
The analysis of the [Li] abundances in these latest samples
does not reveal any appreciable change with
respect to the case of a standard halo metallicity distribution. This 
strenghtens our conclusions, in the sense that they are not critically
dependent on the actual [Fe/H] distribution adopted in the simulations.

\begin{figure}
\centerline{\psfig{figure=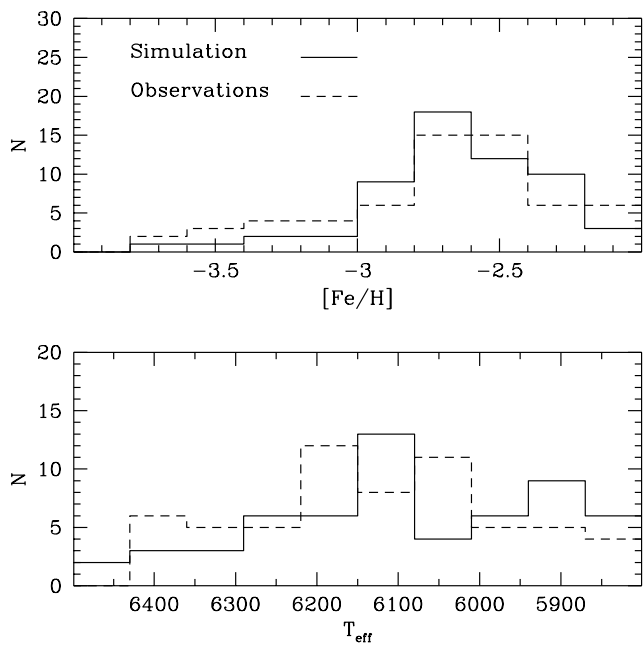,width=9.0cm,clip=}}
\caption[]{ As in the previous figure, but for a synthetic sample 
drawn from a metallicity distribution biased towards the most metal
poor objects (see text for details).}
\protect\label{hist2}
\end{figure}

Before concluding this section, we would like to discuss briefly the
spread of the plateau ($\sigma_{[Li]}^{p}$) obtained in our simulations. 
If we consider all stars in the plateau region
bar the outliers, we obtain a dispersion $\sigma_{[Li]}^{p}$=0.09.
This number compares very well with the values of $\sigma_{[Li]}^{p}$
derived by, e.g., \citen{sfns:96} and \citen{bm:97}, for observational 
errors in [Li] similar to the ones used in our simulations. 
On the other hand, $\sigma_{[Li]}^{p}$=0.09 is definitely larger than
the value given by \citen{rnb:99}. They have reobserved a sample of known
plateau stars (thus avoiding contamination from Li-depleted objects),
claiming average 1$\sigma$ errors by $\sim$30~K in the stellar
temperatures, and by 0.033~dex in the derived [Li] values.
They selected 22 stars around the TO of the field
population, with $T_{\rm eff}$ between
6100$\pm$50~K and 6300$\pm$50~K, [Fe/H] between 
$-$2.5 and $-$3.5, and found that the 1$\sigma$ dispersion of [Li] around an
average value of 2.11 is equal to 0.053~dex (they went on discarding 3 stars
more discrepant from the average to obtain an even smaller dispersion).

We performed a series of MC simulations by drawing samples of 25 stars 
among the samples of plateau stars obtained from the previous simulations. 
We selected the stars in the same [Fe/H] range as \citen{rnb:99}; under
the hypothesis that their hottest stars were stars at the TO, and for avoiding
problems related to zero-point uncertainties in the temperature scale, 
we restricted our sample to stars
with $T_{\rm eff}$ ranging between the hottest value reached in our
simulations and a value 200~K lower. The extremes of this $T_{\rm eff}$ range
are similar to the values adopted by \citen{rnb:99}.
By averaging over 30 of these realizations we obtained an average [Li]=2.17,
with a 1$\sigma$ dispersion of 0.075~dex. This average [Li] value is
lower than the value derived previously; that is easily explained by
the derivative $\Delta [\rm Li]/\Delta [\rm Fe/H]$=0.06 
obtained before, and the
fact that in these latter simulations we are considering only
[Fe/H]$\le -$2.5.
What is probably relevant is that the spread we find is larger
than the value obtained by \citen{rnb:99}, but we would like to recall
the results by \citen{ants:99} which clearly raise the possibility that
current determinations of the plateau dispersion 
(and this would hold also for the larger samples discussed earlier) 
are underestimated appreciably.


\subsection{Field subgiant branch stars}

If one accepts, as a working hypothesis, an age between 13.5 and 14~Gyr
for the most metal-poor field stars, it is
necessary to see if this value satisfies constraints coming from Li
abundances in the SGB phase.

\begin{figure}
\centerline{\psfig{figure=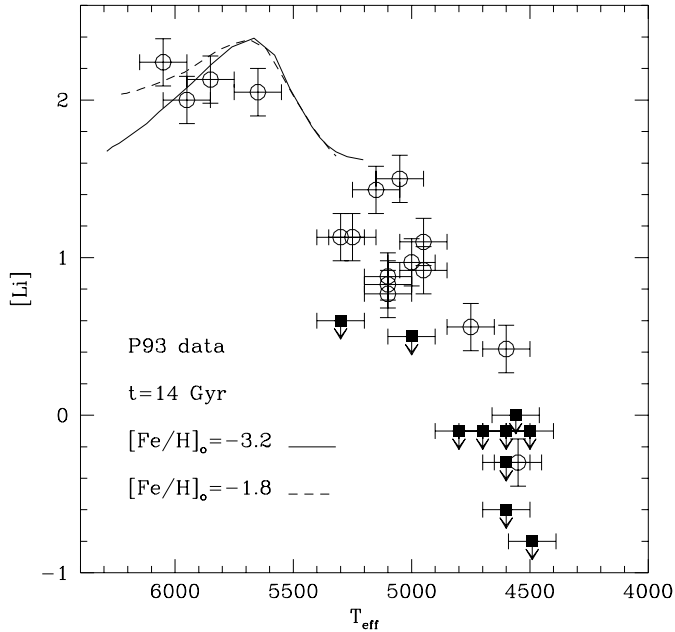,width=9.0cm,clip=}}
\caption[]{Comparison in the [Li]-$T_{\rm eff}$ plane, between
the field subgiant data by \citen{psb:93} (``P93'') for [Fe/H]$\leq
-$2.0, and our isochrones with $t=14$~Gyr for two different initial
metallicities. Squares correspond to upper limits.}
\protect\label{P93}
\end{figure}

We compare our models with observational data 
by \citen{psb:93} for metallicities 
lower than [Fe/H]=$-2$. 
In Fig.~\ref{P93} we show the comparison of the data with the SGB part of our 
isochrones, for an initial $\mathrm{[Fe/H]_o}$ equal to $-$3.2 (solid line) and 
$-$1.8 (dashed line), and $t=14$~Gyr.
Because of the extremely low number of data points on the SGB and
the size of their error bars, we simply compared the position of the
observational data with SGB isochrones bracketing the metallicity
range spanned by the observations.  
Keeping in mind possible systematic errors in the $T_{\rm eff}$ 
scale (which translates also into systematic errors in [Li]), with
respect to the data for the MS, it is 
interesting to notice that the models with diffusion together with our assumed
initial $\mathrm{[Li]_o}$=2.50, are not in contradiction with observations.

The steep decrease of [Li] for decreasing  $T_{\rm eff}$
(equivalently: for increasing luminosity) along
the RGB is neither reproduced by models with diffusion, nor by
standard models (see the discussion in \cite{psb:93}). At
temperatures lower than the endpoint of the isochrones shown in  
Fig.~\ref{P93}, [Li] would stay constant, since the convective
envelope has reached its maximum extension and starts now to recede
slowly due to the outward movement of the H-burning shell.
The further depletion can be explained only by invoking extra deep
mixing (see, e.g., \cite{kra:94}; \cite{dw:96}; \cite{wdc:00}).

\subsection{Globular cluster subgiant branch stars}

\begin{figure}
\centerline{\psfig{figure=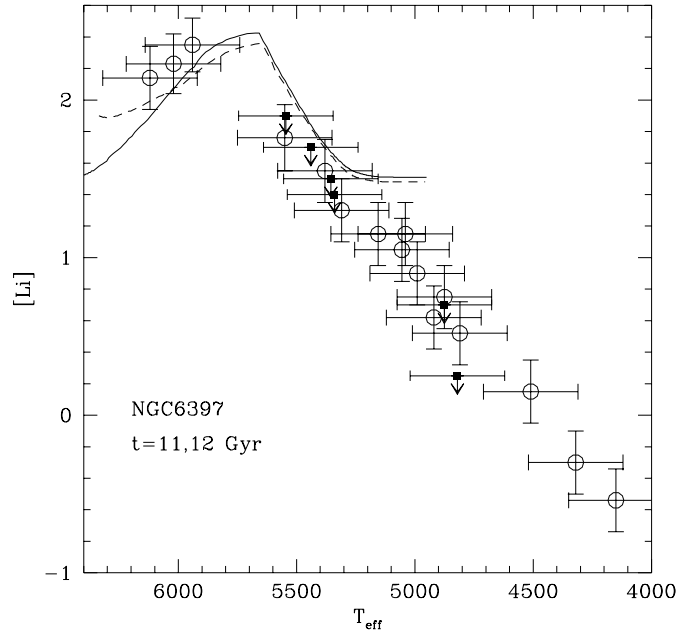,width=9.0cm,clip=}}
\caption[]{Run of [Li] as a function of $T_{\rm eff}$ for SGB and
base of the RGB stars in the globular cluster NGC6397.
Isochrones for ages of 11 and 12~Gyr, and initial
$\mathrm{[Fe/H]_o}$ equal to $-$1.8 are plotted. Symbols are as in Fig.~\ref{P93}.}
\protect\label{N6397}
\end{figure}

\begin{figure}
\centerline{\psfig{figure=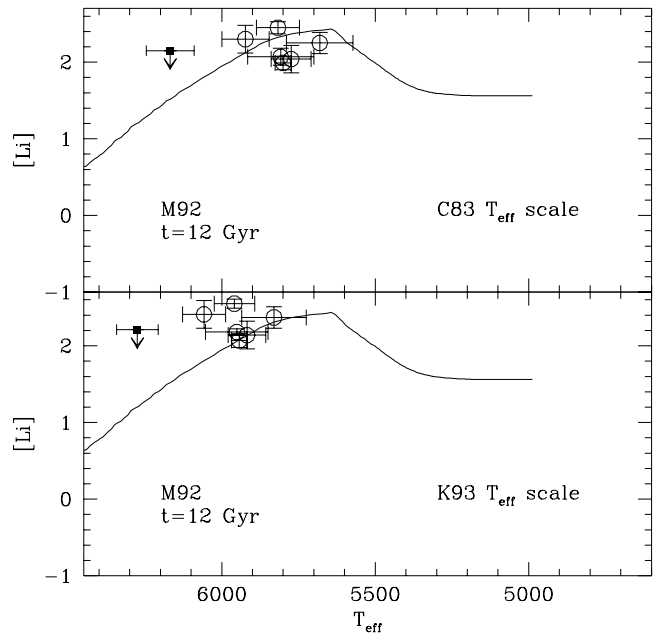,width=9.0cm,clip=}}
\caption[]{As in Fig.~\ref{M92} but for the globular cluster M92. The
theoretical models correspond to ages of 11 and 12~Gyr and an initial
$\mathrm{[Fe/H]_o}$ equal to $-$2.3. The upper panel corresponds to the $T_{\rm
eff}$ scale by C83, the lower panel to the scale by K93.}
\protect\label{M92}
\end{figure}

Figs.~\ref{N6397} and \ref{M92} show a comparison with Li abundances in
globular cluster SGB stars.
In a globular cluster, all stars have the same initial
chemical composition and the same age; in case of NGC6397 we adopted
an initial $\mathrm{[Fe/H]_o}$ equal to $-$1.8, following the results by \citen{cg:97}
for RGB stars belonging to this cluster. This choice is justified by the fact that
at the photosphere of RGB stars the metallicity is restored to the initial
value due to the deepening of the convective envelope, which engulfs
almost all the metals diffused towards the core during the MS phase
(Fig.~\ref{isototal}).  
Ages of 11 and 12~Gyr have been employed, according to the results by \citen{sw:98}.
It is worth noticing that in this case the constraint on the stellar
ages comes from a method completely unrelated to the Li abundance determinations.

The observational data are from \citen{pm:96} 
for SGB stars and \citen{cpa:00} for RGB stars. 
The $T_{\rm eff}$ values by \citen{pm:96} have been reduced by 60~K to
take into account the fact that their assumed 
reddening was 0.01 mag larger than the one used by  \citen{cpa:00};
the corresponding [Li] values consequently have been reduced
according to the derivative $\Delta$[Li]/$\Delta T_{\rm eff}$ provided
in the mentioned papers.

The comparison of the few observed SGB stars with models including
diffusion is satisfactory. 
Notice also the problem along the RGB -- very similar
to the case of field stars -- where models
(standard and diffusive ones) cannot reproduce the trend of [Li] with
repect to $T_{\rm eff}$.

The M92 data displayed in Fig.~\ref{M92} come from \citen{bds:98}; we
considered both temperature scales mentioned in the
paper (C83 is the scale from \cite{c:83}, while K93 is from
\cite{k:93});
this permits us to highlight in a quantitative way the effect of 
$T_{\rm eff}$ scale uncertainties in the [Li] abundances.
It is evident from the figure that, with our selection of initial [Li],
the C83 scale provides a better match to the models. The K93 scale
would imply a higher initial [Li] in order to reproduce better the
observations. In both cases models including diffusion are not in
contradiction with observational data.
The dispersion of [Li] for the sample of stars observed -- about which
a long analysis is presented in \citen{bds:98} -- may be explained by
the fact that (see Fig.~\ref{M92}) the stars are located  
around the peak of the [Li] abundance along the SGB, in a region
where, according to the stellar models,
a small change of $T_{\rm eff}$ 
(within the quoted error bars) produces a sensible change of [Li].
This peak is present only if diffusion on the main sequence has been
effective (Fig.~\ref{isototal}).

It is also interesting to notice that in both cases the point
corresponding to an upper limit detection lies in a region where
models with diffusion predict a lower abundance with respect to the
other stars in the plot.

\subsection{Field stars with $-2.0 <$[Fe/H]$< -$1.5}

\begin{figure}
\centerline{\psfig{figure=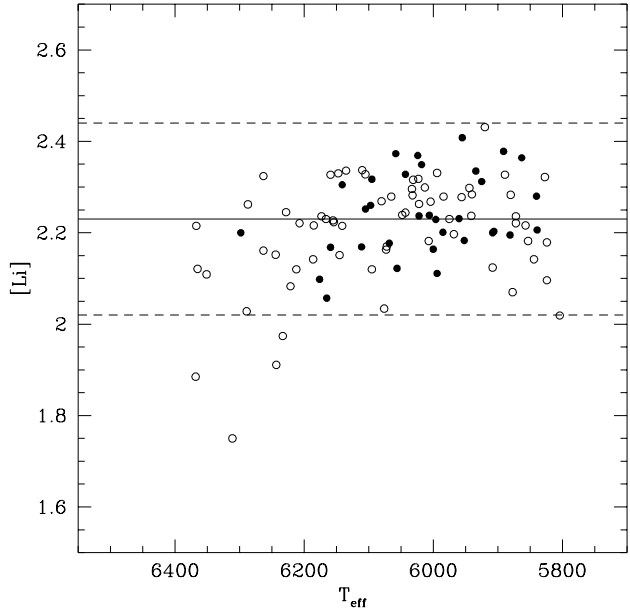,width=8.7cm,clip=}}
\caption[]{ Run of [Li] as a function of 
$T_{\rm eff}$ for stars in one of our MC simulations (see
text for details); open circles are stars 
with [Fe/H]$\le -$2.0, while filled circles are stars
with $-2.0 <$[Fe/H]$< -$1.5.}
\protect\label{metalrichMS}
\end{figure}

Up to now we concentrated on the low-metallicity end of the
Spite-plateau, which, however extends also to higher [Fe/H].
We now consider stars with $-2.0 <$[Fe/H]$< -$1.5, 
in order to see if a plateau can be found also when including 
this metallicity range. We computed a series of models and Li-isochrones 
with initial iron content $\mathrm{[Fe/H]_o}$=$-$1.3 ($Z=0.002$) and then
performed MC simulations selecting this time a sample 
of $\sim$35 plateau stars with $-2.0 <$[Fe/H]$< -$1.5 (which is
approximately the size of the observed sample in this metallicity range).
In Fig.~\ref{metalrichMS} we plot (filled circles) the result of one
simulation (only plateau stars),
together with the result for the stars with 
$\mathrm{[Fe/H]} \le -2.0$ (open circles) from Fig.~\ref{sim2}. 
The solid and dashed lines have the same meaning as in Fig.~\ref{sim2}.
Here we have assumed an age randomly distributed
between 12.5 and 13~Gyr, to simulate a decrease of the average age of
Pop~II field stars for increasing metallicity, above [Fe/H]=$-$2.0.
The average value of [Li] is still 2.23, as in the case of the sample
including only stars with [Fe/H]$\le -$2.0.
The selected age does not play a decisive role. Variations of the
order of $\pm$1~Gyr do not affect appreciably the overall picture. 


\begin{figure}
\centerline{\psfig{figure=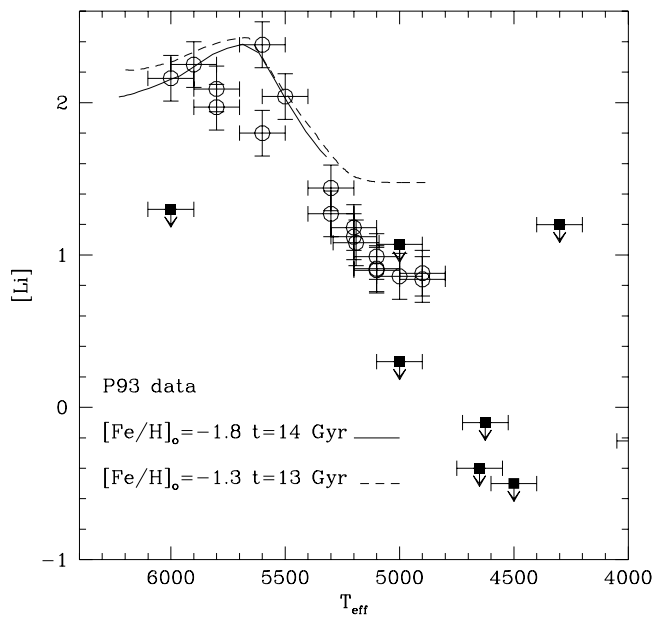,width=8.7cm,clip=}}
\caption[]{Comparison in the [Li]-$T_{\rm eff}$ plane, between
the SGB data by \citen{psb:93} (``P93'') for $-2.0 <$[Fe/H]$< -$1.5, 
and our isochrones with two different initial metallicities and ages. 
Squares correspond to upper limits.}
\protect\label{metalrichSGB}
\end{figure}

Fig.~\ref{metalrichSGB} compares the Li abundances from our models
with the observations by \citen{psb:93} for $-2.0 <$[Fe/H]$< -$1.5.
There appears no inconsistency between theory and
observations besides the fact that also in this metallicity range
there is the well-known 
problem of an additional depletion observed in RGB stars. 

\newpage

\section{Discussion and conclusions}

In the previous section we have compared our stellar models including
uninhibited atomic diffusion with observations of [Li] in metal-poor
([Fe/H]$\le -$2.0) halo field and globular clusters stars.
We have shown that there is no apparent contradiction with
observations, at least with the present
observational data, as long as the age of field subdwarfs is of the
order of 13.5~Gyr or more when [Fe/H]$\le -$2.0.  
This is, of course, not a proof that
diffusion is fully efficient in halo stars (since in general also standard
isochrones can explain the observations; e.g., \cite{vc:95}); 
however, our analysis shows that
it is not justified to discard models with fully efficient diffusion
on the base of current measurements of [Li] in metal-poor stars.

\begin{figure}
\centerline{\psfig{figure=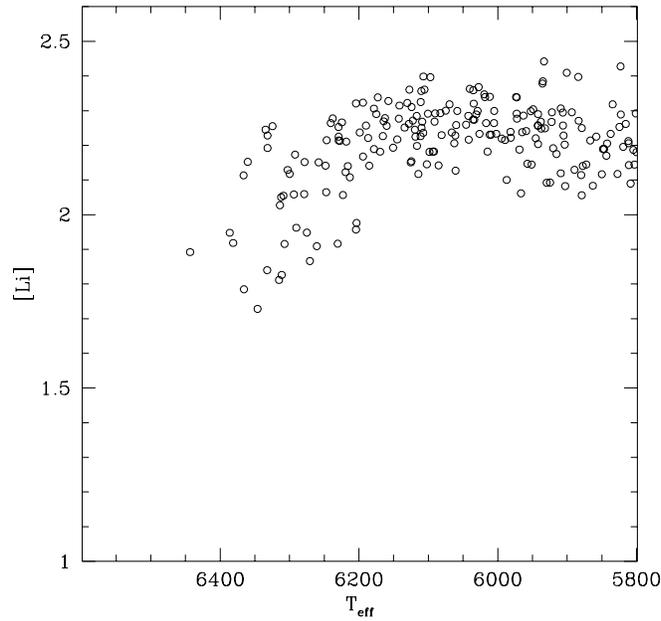,width=8.8cm,clip=}}
\caption[]{Results of a MC simulation with $\sim$200 plateau stars
and [Fe/H]$\le -$2.6 (see text for details). The drop in [Li] at 
the TO due to diffusion is now clearly apparent.}
\protect\label{dropLi}
\end{figure}

In particular, the Li-plateau region is satisfactorily reproduced by
our MC simulations, for the ages discussed before. This is at
odds with conclusions from previous analyses which were based mainly 
on the comparison of an isochrone of a given age and a single value of [Fe/H]
with the moderately small observational samples of stars 
with different metallicities and possibly different ages.
In our MC simulations we could take into account the actual number of
stars with [Li] measurements, their observed metallicity
distribution and the typical observational errors.
The fact that we do not find in the simulations
a clear depletion of [Li] at the hotter end of the plateau is only
partly due to the ages we selected for the stars with [Fe/H]$< -$2.0. 
In fact, lower ages 
(see Fig.~\ref{isoage}) produce a more visible depletion at the TO,
but still, even for ages of about 14~Gyr and taking into account the
observational errors, when $\mathrm{[Fe/H]} \le -2.6$,
a drop at the highest end of the plateau has to be visible, 
provided that a sufficiently high number of stars is observed.
{From} our simulations we derive that [Li] measurements of 
a sample of about 200 plateau stars (about 40-50 of them at the hottest
end of the plateau)
with [Fe/H]$\le -$2.6 are needed
in order to clearly detect this drop, keeping unchanged 
the observational errors (see Fig.~\ref{dropLi}).

For the more metal-rich objects of section~3.4 with [Fe/H] above 
$-$2.0 an increase of the observed sample size would not change 
the plateau morphology, since at these metallicities
there is no abrupt drop of [Li] at the TO region, but an almost 
constant value of [Li] all along the plateau region.

\begin{figure}
\centerline{\psfig{figure=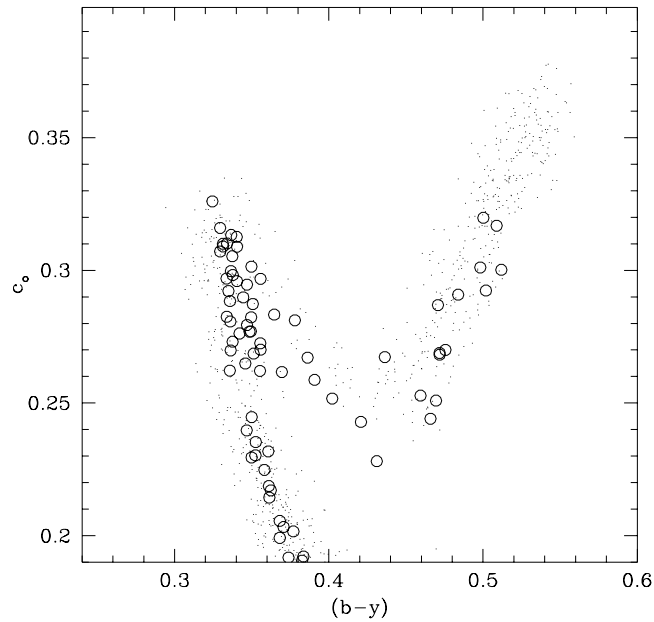,width=8.8cm,clip=}}
\caption[]{$\rm c_{0}$--(b-y) diagram for a simulated 
sample of field stars (open circles), and for a synthetic globular
cluster with [Fe/H]=$-$2.3 and $t=12$~Gyr 
(dots -- see text for details).}
\protect\label{FiCl}
\end{figure}

In our simulations we have treated the age of field stars as a free
parameter, requiring that it is not too far from current estimates of globular
clusters ages, and that with the actual samples of observed stars
the plateau is well reproduced.
Is an age of 13.5--14.0~Gyr for field Population II stars of 
$\mathrm{[Fe/H]} \le -2.0$ compatible with their CMD?
It is hard to answer this question with confidence. Due to the 
uncertainties in current colour--transformations (see, e.g., \cite{ws:99})
the best thing to do, in our opinion, is to determine the
location of field subdwarfs in a CMD or colour--colour diagram, in
relation to 
data for a globular cluster of known age and similar metallicity, as
predicted by theory, and then check its consistency with observations.

{From} the observational point of view \citen{g:99} has shown how, in
Stroemgren colours, M92 stars overlap with the metal-poor field objects
([Fe/H]$\le -$2.5) in the sample by \citen{snp:96}.
In Fig.~\ref{FiCl} we show the theoretical counterpart of the
comparison performed by \citen{g:99}; we used our stellar models
with diffusion, coupled to the colour transformations provided by Kurucz
(http://cfaku5.harvard.edu/). 
The open circles represent
a MC realization of a sample of field stars with age between 13.5 and 14~Gyr, 
an actual [Fe/H]$\le -$2.5, observed metallicity distribution according to \citen{rn:91},
1$\sigma$ observational errors by 0.010 mag in (b-y) and
$\rm c_{0}$, and 0.20~dex in [Fe/H]; this sample simulates 
the sample by \citen{snp:96}. Dots are a MC realization of a
sample of stars with the same initial metallicity ($\mathrm{[Fe/H]_o}$=$-$2.3),
an age of 12~Gyr, and with 1$\sigma$
observational errors by 0.010 mag in (b-y) and $\rm c_{0}$.  
This sample simulates the data of M92.

It is clear from the figure that the two samples nicely overlap, in
agreement with observations (see Fig.~1b in
\cite{g:99}). Even if the constraint put by this 
comparison is not too stringent, in the sense that a variation of the
relative age between cluster and field by 1-2~Gyr would not affect
this comparison appreciably, it is still 
a nice consistency check for the results derived from the shape of the
lithium plateau.


It has been mentioned several times in the literature that
$^6\mathrm{Li}$, which is destroyed by proton-captures at 
even lower temperatures than $^7\mathrm{Li}$ 
($2\cdot 10^6\,\mathrm{K}$), would be an additional and
strong indicator of $^7\mathrm{Li}$ destruction (see, for example,
\cite{vcca:99}). 
What observations provide is the sum $^7\mathrm{Li}$+$^6\mathrm{Li}$.
Disentangling the abundance of one isotope from the other one is a
very difficult task;
it is based on fitting the observed profile of the Li I blend 
with theoretical profiles calculated from model stellar atmospheres 
characterized by different $^6\mathrm{Li}$/$^7\mathrm{Li}$ ratios
(see, e.g., \cite{htr:99}).
Until now all (very few) measured Li-isotope ratios (e.g.\
\cite{nlps:99}; \cite{htr:99}; \cite{navd:00}) are either upper limits
only or are so high ($^6\mathrm{Li}/(^7\mathrm{Li}+^6\mathrm{Li})
\approx 0.05$) that
they exceed limits from Big Bang Nucleosynthesis by orders of
magnitude, such that galactic chemical evolution (\cite{vcca:99}) has
to be taken into account to explain them. {From} this point on, of course,
all conclusions about stellar depletion of lithium isotopes are
subject to the acceptance of these chemical evolution models. 
The quoted papers arrive at conclusions both excluding
(\cite{vcca:99}) and requiring stellar depletion (\cite{htr:99}). 
We therefore conclude that presently $^6\mathrm{Li}$ does not provide
additional conclusive information about diffusive processes in
low-mass stars. We note that along the MS diffusion alone
would only mildly change the isotope ratio, anyway, with an increase
of the initial $^6\mathrm{Li}/(^7\mathrm{Li}+^6\mathrm{Li})$ value 
by at most 40\%.

In principle, useful constraints on diffusion in low-mass stars could
come from spectroscopy along globular cluster CMDs, as the one shown
in Fig.~\ref{N6397} for NGC6397. Recently, \citen{gbbc:00} have
presented VLT/UVES results about element abundances in stars at
the TO and at the RGB base in NGC6397 and NGC6752. The results,
however, are far from being conclusive. In NGC6752 large star-to-star
scatter in light element abundances is present, indicating additional
stellar processes. This applies also to lithium (Gratton, private
communication). In NGC6397 rather constant abundance values are found,
and in particular [Fe/H] appears to be exactly the same in both groups
of stars, a fact which would strongly disfavour diffusion. However, in
a paper about the same cluster, \citen{cpa:00} discussed the
uncertainty in metal abundance determinations, finding an iron {\em
decrease} by 0.2~dex when moving along the RGB, if the temperature
scale of \citen{gcc:96} is used. Also \citen{gbbc:00} find a
metallicity at the TO and base of the RGB that is  0.2~dex lower than
the \citen{cg:97} value.  
Since the depletion of iron due to diffusion that our models
predict for NGC6397 is of the same order, one should remain cautious about the
reality of constant [Fe/H] values determined. More work is needed to
clarify this. 

\citen{rcbb:01} have very recently presented results about [Fe/H]
abundances for stars from the RGB tip to the TO of the metal rich 
globular cluster M71 ([Fe/H]$\approx$0.70). They found 
that the iron abundances are 
approximately constant along these evolutionary phases. However, as we
verified with appropriate calculations, in this metallicity range the
maximum depletion at the TO, for ages of the order of 10 Gyr 
(\cite{sw:98}) is 
$\sim$0.15 dex. Such a small depletion is not in contradiction with the
values provided by \citen{rcbb:01} in their Table 4.


Assuming that diffusion indeed has taken place in Spite-plateau stars,
our results have significant bearings for the question
concerning the cosmological baryon density. 
We reproduce the observed
Spite-plateau [Li]-abundance when assuming a primordial value of
$\mathrm{[Li]} = 2.5$. According to standard Big Bang Nucleosynthesis
(see, for example, the recent calculations by \cite{vfcc:00}), this
corresponds to a baryon-to-photon ratio $\eta$ of $5.1\cdot 10^{-10}$
(see Fig.~4 of the quoted paper), which is in almost perfect agreement
with the low deuterium value  ($\mathrm{D/H}=3\ldots 4\cdot
10^{-5}$) by \citen{thsl:00} and the high $^4$He abundance of 0.244 by
\citen{it:98}. Since $\eta$ is related to the baryonic density via
$\eta = 273\cdot 10^{-10} \Omega_\mathrm{B} h^2$ (with $h$ being the
Hubble constant in units of $100\, \mathrm{km}\,\mathrm{s}^{-1}\,
\mathrm{Mpc}^{-1}$), this results in $\Omega_\mathrm{B} = 1.87\cdot
10^{-2} h^{-2}$. Taking $h=0.71$ (\cite{mhfkf:00}) yields
$\Omega_\mathrm{B} = 0.037$, almost consistent with the value
inferred from the BOOMERANG and MAXIMA-1 experiments
(\cite{jaffetal:00}; \cite{babb:00}).

Our derived cosmological baryon density is under the assumption
of no significant galactic production of $^7\mathrm{Li}$. 
$^6\mathrm{Li}$ measurements can in principle
be used also to put useful constraints on the amount of $^7\mathrm{Li}$
produced by galactic chemical evolution, but the available 
measurements of the abundance ratio of these two isotopes (see also
previous discussion) do not provide yet a consistent picture. 
As an example, the value $^6\mathrm{Li}/(^7\mathrm{Li}+^6\mathrm{Li})
\approx 0.05$ for the metal poor 
([Fe/H]$\approx -$2.3) TO star HD 84937 points to a
non-negligible contribution
of $^6\mathrm{Li}$ -- and therefore also $^7\mathrm{Li}$ -- from
cosmic ray production. However, measurements of the same quantity for
HD 218502, a star with similar metallicity and
evolutionary status, provide an upper limit of only 0.02 
and a best fit value equal to zero (see \cite{htr:99}), consistent with
no contribution from galactic chemical evolution.
If in the future $^6\mathrm{Li}$ measurements will consistently indicate
a significant influence of galactic production of $^7\mathrm{Li}$
(of order 0.1 dex or more), one will have to reconsider our
conclusions about the cosmological baryon density.

One could in principle worry about the fact that our adopted initial [Li] 
abundance implies a primordial $Y\sim$0.24, while in our
models we used $Y=0.23$. However, as we have verified with appropriate
computations, the effect of using $Y=0.24$ as the initial He abundance
for our models is just an age shift. More in detail,
the [Li] abundance as a function of $T_{\rm eff}$
along an isochrone of fixed initial $\mathrm{[Fe/H]_o}$ and age 
computed with an initial $Y=0.23$ would be equal to the abundances of an isochrone
with the same initial $\mathrm{[Fe/H]_o}$ but an age lower by 1~Gyr, if an
initial $Y=0.24$ is used. The same holds for the [Fe/H] values along an isochrone.
Our conclusions about the Li-plateau are therefore not affected at all, 
apart from the fact that in
this case the minimum age required to reproduce the observed
plateau would be 12.5~Gyr.


We finally mention that our treatment of
diffusion does not take into account the effect of radiative
levitation. The diffusive velocities by
\citen{tbl:94}, which we are using, might be overestimated 
by 25\% on average, at least in the case of the Sun, 
according to \citen{trmir:98}. It is plausible,
but not certain, that a similar overestimate might happen for
metal-poor stars. We neither included mass loss in the calculations,
which is also known to reduce the effect of diffusion. 
For mass loss rates compatible with current observations 
the surface abundances depletion is not affected appreciably, but
strong stellar wind on the MS may inhibit diffusion completely
(\cite{vc:95}). 
Due to the neglect of both radiative levitation
and mass loss, our models probably overestimate the depletion to some 
degree, even if we cannot quantify this up to now.

We can conclude that the Spite plateau, given present 
day observational sample sizes, can be reproduced by models including
atomic diffusion and is therefore (alone) no strong argument against
the presence of sedimentation in low-mass metal-poor stars. We also
recall that the motivation for this work was the fact that
sedimentation in the {\em cores} of such stars (in globular clusters)
affects their evolutionary speed and thereby age determinations. What
the Spite plateau represents, however, is related to diffusion from the
{\em envelopes} of field stars. Both things have the same physical
origin, but conclusions regarding the latter might not be applied
straightforwardly to the former.


\begin{acknowledgements}
A.W.\ is grateful for the hospitality at the Institute for Advanced Study 
and Princeton Observatory and for a Fulbright fellowship which allowed
visiting both places. Helpful discussions with J.~Truran, L.M.~Hobbs,
R.~Cayrel, C.~Deliyannis, K.~Jedamzik, and F.~D'Antona are acknowledged. 
\end{acknowledgements}
%

\end{document}